\documentclass[12pt,thmsa]{article}

\usepackage{amsfonts}
\usepackage{graphicx}
\usepackage{epsfig}
\usepackage{graphics}
\makeatletter
\newcommand{\row}[1]{\mathord{\buildrel{\lower3pt\hbox{$\scriptscriptstyle\rightarrow$}}\over #1}}

\newcommand{\dyadic}[1]{\mathord{\dyadic@rrow{#1}}}
\newcommand{\dyadic@rrow}[1]{
\begin{picture}(12,12)(-1,0)
\put(-2,12){\makebox(0,0)[t]{$\scriptscriptstyle\downarrow$}}
\put(-2,12){\makebox(0,0)[l]{$\scriptscriptstyle\longrightarrow$}}
\put(5,0){\makebox(0,0)[b]{$#1$}}
\end{picture}
}
\newcommand{\bra}[1]{\bigl\langle #1 \bigr|}
\newcommand{\ket}[1]{\bigl| #1 \bigr\rangle}

\newcommand{\partr}[1]{^{\mathsf{T}_{\!#1}}}
\input{tcilatex}

\begin{document}
\begin{center}
{\large  Long-lived entanglement with Pulsed-driven  initially entangled qubit pair}\\
\vspace{0.2cm}
{\small N. Metwally$^{1,2}$, H. A. Batarfi$^{3}$ and S. S.Hassan$^{2}$\\
\vspace{0.2cm} $^{1}$Department of Mathematics, Faculty of
Science, Aswan University, Aswan,
Egypt.}\\
{\small $^{2}$Department of Mathematics, College of Science, University of Bahrain,\\
P. O. Box 32038 Kingdom of Bahrain \\

 $^{3}$Department of Mathematics, College of Science, King AbdelAziz
 University,\\
P. O. Box 41101 Jeddah 21521, Kingdom of Saudi Arabia}\\
nmetwally@gmail.com$^*$, hbatarfi@kau.edu.sa,
shoukryhassan@hotmail.com

\end{center}

\begin{abstract}
The entanglement of different classes of initially entangled qubit
pair  is investigated in the presence of short laser pulses of
rectangular and exponential shapes with either one or both qubits
are excited. For the rectangular pulse, the detuning parameter
 protects the entanglement from degradation
 and controls its upper and lower bounds.
We show that the upper bounds of entanglement decrease if both
particles are excited, but the lower bounds are much better if
only one particle is excited.
 The phenomena of  entanglement degradation,
sudden- death and -birth  are shown for small initial time. Long
-lived entanglement behavior, but with a variable degree, appears
in the presence of rectangular pulse. However,  in the case of the
exponential pulse one obtains an invariably long- lived
entanglement. For the combined case (rectangular $+$ exponential)
, we show that one can generate a long-lived entanglement with
small variable by increasing the strength  of  the rectangular
pulse area and decreasing the Rabi frequency of the exponential
pulse.
\end{abstract}

 {\bf Keywords:}  Pulsed- driven qubits, Entanglement,

\section{Introduction}

One important research area within the context of quantum
information science is  investigating the dynamics of travelling
states that are utilized to perform  quantum information tasks as
quantum teleportation  and coding \cite{Barnett}. These states
could be subject to  noise channels \cite{Ping,Nasser0} or
dissipative environment \cite{Ikram} and consequent interactions
take place. These interactions may lead to undesirable loss of
quantum entanglement (eq.  decay and sudden- death ) that is
essential for these states to perform in quantum information
processing and computation \cite{Gab,Zhao}.

Non-dissipative single qubits when exposed to laser pulses have
their energy levels splitting dependent on the pulse strength and
its off-resonance  parameter. Fourier and wavelet spectral
features of the fluorescent radiation of such energy splitting
have been investigated for various shapes of laser pulses (
\cite{Rod}-\cite{Bata1}). Likewise, travelling quantum states that
are exposed to pulsed lasers suffer a change in its entangled
properties. Information transfer and orthogonality speed of a
single qubit driven by  a rectangular pulse has been recently
investigated by us \cite{Nasser1}. Recently it has been shown that
quantum correlations can be enhanced and protected by applying a
train of instantaneous pulses (so called bang-bang pulses) on a
two qubit system \cite{Hang}. Therefore, it is desirable  to study
the behavior of systems of a single qubit and two qubits when
exposed to different shape of laser pulses.

In the present work, we investigate  entangled properties  of a
class of generalized Werner state,   including   partial and
maximum entangled states, interacting  with two types of pulses,
namely, rectangular and exponential pulses. Time evoluation of the
initial state is obtained analytically for different cases and the
 the amount of the survival entanglement is quantified and computed accordingly.

The paper is organized as follows. In Sec.2, we present the model
and its  exact operator  solution. The dynamics of entanglement is
quantified in  Sec.3, where we use the negativity as a measure of
entanglement. The effect of the detuning parameter and the Rabi-
frequency on the entanglement is investigated. Degree of
entanglement in the three cases of   pulse excitation of one or
both particles, namely, rectangular, exponential and combination
of them, is investigated in Sec.(3-5), respectively. A summary is
given in Sec. 6.

\section{The suggested Model}

Assume that a source supplies two users, Alice and Bob, with  a
partially entangled (generalized Werner) state \cite{Englert}
defined by,
\begin{equation}\label{ini}
\rho=\frac{1}{4}(1+c_{xx}\sigma^{(a)}_x\sigma_x^{(b)}+c_{yy}\sigma_y^{(a1)}\sigma_y^{(b)}+c_{zz}\sigma_z^{(a)}\tau_z^{(b)}).
\end{equation}
In (\ref{ini}) $a,b$ stand for Alice and Bob's qubit,
respectively. It assumed that during the transformation from the
source to the users' positions, each qubit interacts with  a
different pulse. The full Hamiltonian of this system, within the
rotating wave approximations (in units of $\hbar=1$) is given by
\cite{shukry},
\begin{equation}
\hat{H}^{(j)}=\omega_a\hat{S}_z^{(j)}+\frac{\Omega^{(j)}(t)}{2}\Bigl\{\hat{S}_{+}^{(j)}
e^{-i\omega^{(j)}_l t}+ c.c)\Bigr\},
\end{equation}
where $i=a$ (Alice qubit) and $i=b$ (Bob qubit), the
spin-$\frac{1}{2}$ operators $\hat{S_{\pm}^{(i)}},\hat{S_z^{(j)}}$
obey the $Su(2)$ algebra,
\begin{equation}
[\hat{S}_{+}^{(j)}, \hat{S}_{-}^{(j)}]=2\hat{S}_{z}^{(j)},\quad
[\hat{S}_{z}^{(j)},\hat{S}_{\pm}^{(j)}]=\pm\hat{S}_{\pm}^{(j)}.
\end{equation}
The parameter $\Omega^{(j)}(t)=\Omega_0^{(j)} f^{(j)}(t)$ where
$\Omega^{(j)}$ represents the real Rabi frequency associated with
the laser pulse  and $f^{(j)}(t)$ is a dimensionless parameter
describes the pulse shape. For the rectangular pulse this function
is defined as
\begin{equation}
 f^{(j)}(t) = \left\{ \begin{array}{ll}
1&  \textrm{ for  $t\in[0,T]$}\\
0 & \quad\textrm{otherwise},\\
\end{array} \right.
\label{equation6}
\end{equation}
where the pulse duration $T$ is much shorter than the lifetime of
the atomic upper state, hence atomic damping can be discarded. For
the exponential pulse of widith $\gamma_p$ the function
$f^{(i)}(t)$  is defined as \cite{Bata1}
\begin{equation}
 f^{(j)}(t) = \left\{ \begin{array}{ll}
e^{-\gamma_p t}&  \textrm{ for  $t\geq 0$}\\
0 & \quad\textrm{for $t<0$}.\\
\end{array} \right.
\label{equation6}
\end{equation}

To solve the system which is given by the equations (1,2), we
introduce operators.
\begin{equation}
\hat{\sigma}_{\pm}^{(j)}(t)=\hat{S}^{(j)}_{\pm}(t)e^{\mp
i\omega^{(j)}_l t}, ~\quad
\hat{\sigma}^{(j)}_z(t)\equiv\hat{S}^{(j)}_z(t),
\end{equation}
in  the rotating frame where  the $\hat{\sigma}$ operators obey
the same algebraic form of Eq.(3). The time evolution of the
operators $\hat{\sigma}_{\pm,z}(t)$ according to (2) are given by,
\begin{eqnarray}
\hat{\sigma}^{(j)}_{x}(t)&=&
\mathcal{A}^{(j)}_{x}(t)\hat\sigma_x(0)+
\mathcal{A}^{(j)}_{y}(t)\hat\sigma_y(0)+\mathcal{A}^{(j)}_{z}(t)\hat\sigma_z(0),
\nonumber\\
\hat{\sigma}^{(j)}_{y}(t)&=&
\mathcal{B}^{(j)}_{x}(t)\hat\sigma_x(0)+
\mathcal{B}^{(j)}_{y}(t)\hat\sigma_y(0)+\mathcal{B}^{(j)}_{z}(t)\hat\sigma_z(0),
\nonumber\\
\hat{\sigma}^{(j)}_{z}(t)&=&\mathcal{D}^{(j)}_{x}(t)\hat\sigma_x(0)+
\mathcal{D}^{(j)}_{y}(t)\hat\sigma_y(0)+\mathcal{D}^{(j)}_{z}(t)\hat\sigma_z(0),
\end{eqnarray}
where, the $c$-number time -dependent functions
$\mathcal{A}_i(t),\mathcal{B}_i(t)$ and $\mathcal{D}_i(t)$ are
given by
\begin{eqnarray}
\mathcal{A}^{(j)}_{x}(t)&=&\frac{1}{2}\Bigl[\mathcal{C}^{(j)}_{+}(t)+
\mathcal{C}^{(j)}_{-}(t)+c.c.\Bigr],\quad
\mathcal{A}^{(j)}_{y}(t)=\frac{i}{2}\Bigl[\mathcal{C}^{(j)}_{+}(t)-
\mathcal{C}^{(j)}_{-}(t)-c.c.\Bigr],\quad
\nonumber\\
\mathcal{A}^{(j)}_{z}(t)&=&\frac{1}{2}(\mathcal{C}^{(j)}_{z}(t)+c.c)
\nonumber\\
\mathcal{B}^{(j)}_{x}(t)&=&-\frac{i}{2}\Bigl[(\mathcal{C}^{(j)}_{+}(t)+\mathcal{C}^{(j)}_{-}(t)-c.c.)\Bigr]
,\quad \mathcal{B}^{(j)}_{y}(t)=i\mathcal {B}_x^{(j)}(t),\quad
\mathcal{B}^{(j)}_{z}(t)=-i\mathcal{A}^{(j)}_{z}(t)
\nonumber\\
\end{eqnarray}
 The  expressions  for the coefficients  $\mathcal{C}^{(j)}_{\pm}, \mathcal{C}^{(j)}_{z},D^{(j)}_x,D^{(j)}_y$ and $D^{(j)}_z$   will be
given  below  according to the type of the applied pulse. By using
these results, the time evolution of the initial state (1) is
given by
\begin{eqnarray}
\rho_{ab}(t)&=&\frac{1}{4}\Bigl[1+\tilde{C}_{xx}\sigma^{(a)}_x\sigma_x^{(b)}+\tilde{C}_{xy}\sigma^{(a)}_x\sigma_y^{(b)}
+\tilde{C}_{xz}\sigma^{(a)}_x\sigma_z^{(b)}
+\tilde{C}_{yx}\sigma^{(a)}_y\sigma_x^{(b)} \nonumber\\
&+& \tilde{C}_{yy}\sigma^{(a)}_y\sigma_y^{(b)}
+\tilde{C}_{yz}\sigma^{(a)}_x\sigma_z^{(b)}+
\tilde{C}_{zx}\sigma^{(a)}_z\sigma_x^{(b)}+\tilde{C}_{zy}\sigma^{(a)}_z\sigma_y^{(b)}
+\tilde{C}_{zz}\sigma^{(a)}_z\sigma_z^{(b)}\Bigr],
\end{eqnarray}
where, the nine coefficients $\tilde C_{kl}(t)$; $k,l=x,y,z$ are
given by,
\begin{equation}
\tilde
C_{kl}(t)=\mathcal{A}^{(1)}_{k}(t)\mathcal{A}^{(2)}_{l}(t)c_{xx}(0)+
\mathcal{B}^{(1)}_{k}(t)\mathcal{B}^{(2)}_{l}(t)c_{yy}(0)+
\mathcal{D}^{(1)}_{k}(t)\mathcal{D}^{(2)}_{l}(t)c_{zz}(0)
\end{equation}
In the following two sections, we  evaluate the time evolution of
the initial state described by Eq.(1) for different types of
pulses. we investigate the dynamics of entanglement of different
classes of initial states. If we set
$c_{xx}(0)=c_{yy}(0)=c_{zz}(0)=-1$, one gets the maximum entangled
single state \cite{Nasser}. For the Werner state,
$c_{xx}(0)=c_{yy}(0)=c_{zz}(0)=x$, which represents  partially
entangled states for $-\frac{1}{3}\leq x\leq \frac{1}{3}$ and
seperable otherwise.

As for  the dynamics of entanglement, we use the negativity as a
measure of entanglement. This measure has been introduced by K.
Zyczkowski et. al \cite{Zyc} and  states that if the eigenvalues
of the partial transpose of $\rho_{ab}(t)$, namely,
$\rho_{ab}^{\partr 2}(t) $ are given by $\mu_i; i = 1, 2, 3, 4,$
then the degree of entanglement (DOE) is given by,
\begin{equation}
\mathcal{E}(t)=\sum_{i=1}^{4}{\mu_i}-1,
\end{equation}
where, $0<\mathcal{E}(t)<1$.

\section{ Rectangular  Pulse Case}
In this section we consider the effect of the rectangular pulse
(4) on the degree of entanglement. In this case, the coefficients
$\mathcal{C}^{(j)}_{\pm}(t), \mathcal{C}^{(i)}_{z}(t),
D^{(j)}_x(t) $, $D^{(j)}_y(t)$ and $D^{(j)}_z(t)$ are given by,
(cf \cite{shukry}),
\begin{eqnarray}
\mathcal{C}^{(j)}_{+}(t)&=&\frac{1}{2}\Bigl\{\left(\frac{\Omega^{(j)}}{\Omega^{(j)}_1}\right)^2+
\left(\frac{\Delta^{2(j)}+\Omega_1^{2(j)}}{\Omega_1^{2(j)}}\right)\cos(\Omega_1^{(j)}
t)\Bigr\}
+i\frac{\Delta^{(j)}}{\Omega_1^{(j)}}\sin(\Omega_1^{(j)}t)
 \nonumber\\
\mathcal{C}^{(j)}_{-}(t)&=&\frac{1}{2}\left(\frac{\Omega^{(j)}}{\Omega^{(j)}_1}\right)^2(1-\cos(\Omega_1^{(j))}
t)),
 \nonumber\\
\mathcal{C}^{(j)}_{z}(t)&=&\frac{\Delta^{(j)}\Omega^{(j)}}{(\Omega^{(j)}_1)^2}\Bigl(1-\cos(\Omega^{(j)}_1
t)\Bigr)-i\frac{\Omega^{(J)}}{\Omega_1^{(j)}}\sin(\Omega_1^{(j)}
t)
\nonumber\\
D^{(j)}_x(t)&=&\frac{\Delta^{(j)}\Omega^{(j)}}{\Omega_1^{2(j)}}\Bigl(1-\cos(\Omega_1^{(j)}t)\Bigr),\quad
D^{(j)}_y=\frac{\Omega^{(j)}}{\Omega_1^{(j)}}\sin(\Omega_1^{(j)}t),
\nonumber\\
D^{(j)}_z(t)&=&\Bigl(\frac{\Omega^{(j)}}{\Omega_1^{(j)}}\Bigr)^2\Bigl[\cos(\Omega_1^{(j)}t)+
\Bigl(\frac{\Delta^{(j)}}{\Omega^{(j)}}\Bigr)^2\Bigr],
\end{eqnarray}
where $\Omega_1^{(j)}=\sqrt{(\Omega^{(j)}_0)^2+(\Delta^{(j)})^2}$,
$\Delta^{(j)}=\omega^{(j)}-\omega^{(j)}_l$.\\
Using Eqs.(9) and (12), one obtains the time evoluation of the
initial density operator(1) in the presence of the local
rectangular pulse. In this context, we investigate the effect of
the detuning parameter $\Delta^{(j)}$ and the  Rabi frequency
$\Omega^{(j)}$ on the dynamics of the entanglement.

We  consider three different initial states of the two non-
interacting qubits: \\(i) Maximum entangled state which is
characteristed by $c_{xx}(0)=c_{yy}(0)=c_{zz}(0)=-1$. This state
is defined as Bell state (singled state),
$\rho_{\psi^-}=\ket{\psi^-}\bra{\psi^-}$
where $\ket{\psi^-}=\frac{1}{\sqrt{2}}(\ket{01}-\ket{10})$,\\
  (ii) Werner state, where we assume that $c_{xx}(0)=c_{yy}(0)=c_{zz}(0)=-0.9$,
which represents a class of partially entangled states, and \\
 (iii) Generalized Werner state, or $x-state$, where we assume that
$c_{xx}(0)=-0.9,c_{yy}(0)=-0.8,$ and $c_{zz}(0)=-0.6$.

Having prepared the two qubit initially in any of the three
aforementioned entangled states, we consider next two cases:

\subsection{One driven-qubit}

Here, we assume that only one qubit is driven by a rectangular
pulse. This means we excite one qubit locally and then look
globally for the entanglement of the whole system. We calculate
the entanglement parameter $\mathcal{E}(T)$ at the end of the
pulse duration $t=T$ and take the pulse area parameter $\Omega
T/2\pi=n$ and define the normalized detuning
$\Delta'^{(j)}=\Delta^{(j)}/\Omega$. At exact resonance
$\Delta'^{(1)}=0$ (Fig.(1a)), $\mathcal{E}(T)$ keeps its initial
values for integer values of (n), whilst for non-integer values of
$(n)$ it is relatively reduced. For non-zero detuning ,
$\Delta'^{(1)}\neq 0$, Fig.(1b) shows that with increasing pulse
area parameter $n=20$, $\mathcal{E}(T)$ approaches zero value with
the initial generalized  Werner state i.e., the less entangled
initial state turns into separable state, while the maximum
initial state (Bell-state) keeps the entanglement to non-zero
lower bound. Further increase  of $\Delta'^{(1)}$ causes increase
in the lower bounds of entanglement increases for all initial
states. Also, $\mathcal{E}(T)$  shows its periodic behaviour with
further increase of (n).

\begin{figure}[t!]
\begin{center}
\includegraphics[width=18pc,height=12pc]{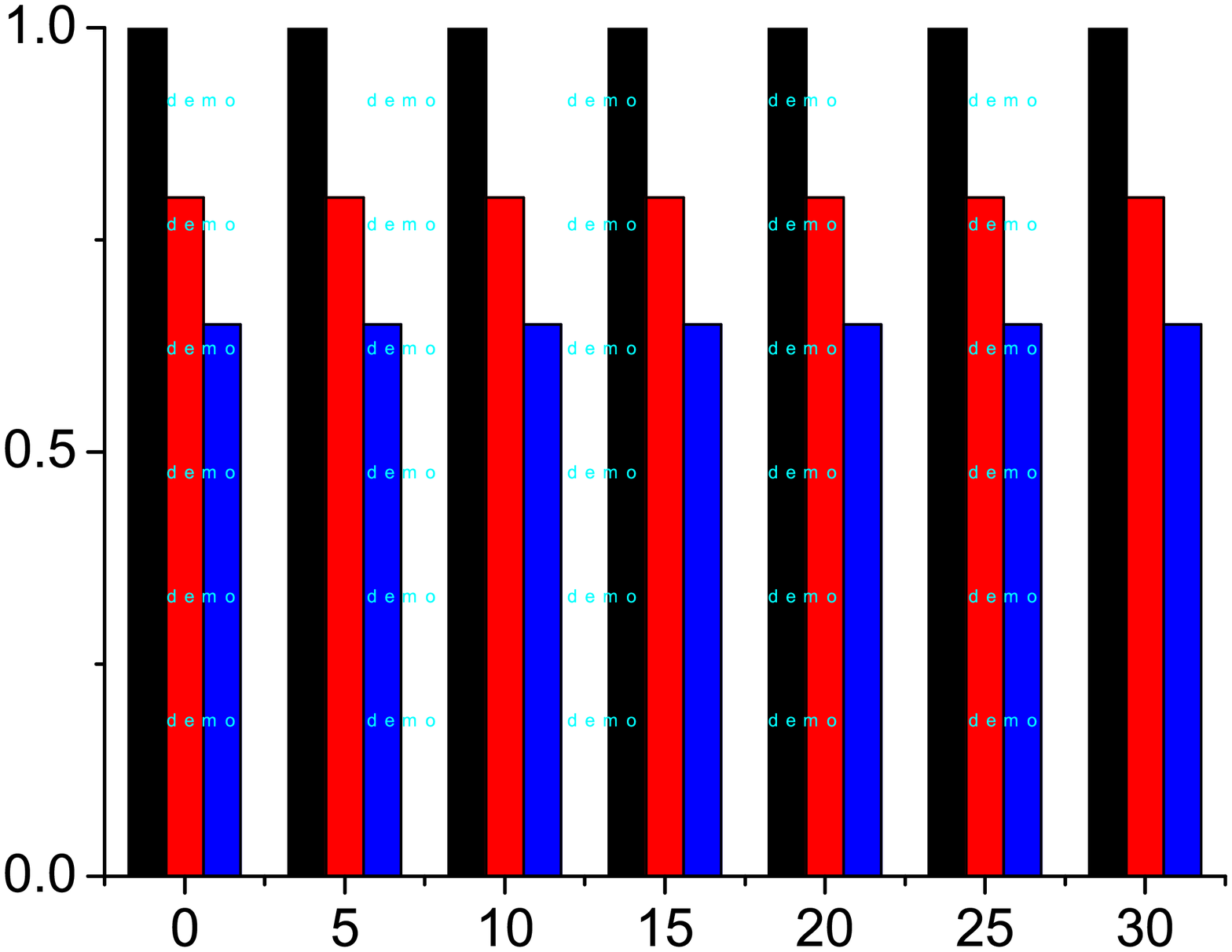}
\put(-20,120){${(a)}$}
 \put(-215,70){$\mathcal{E}(T)$}
 \put(-100,1){${n}$}
\includegraphics[width=18pc,height=12pc]{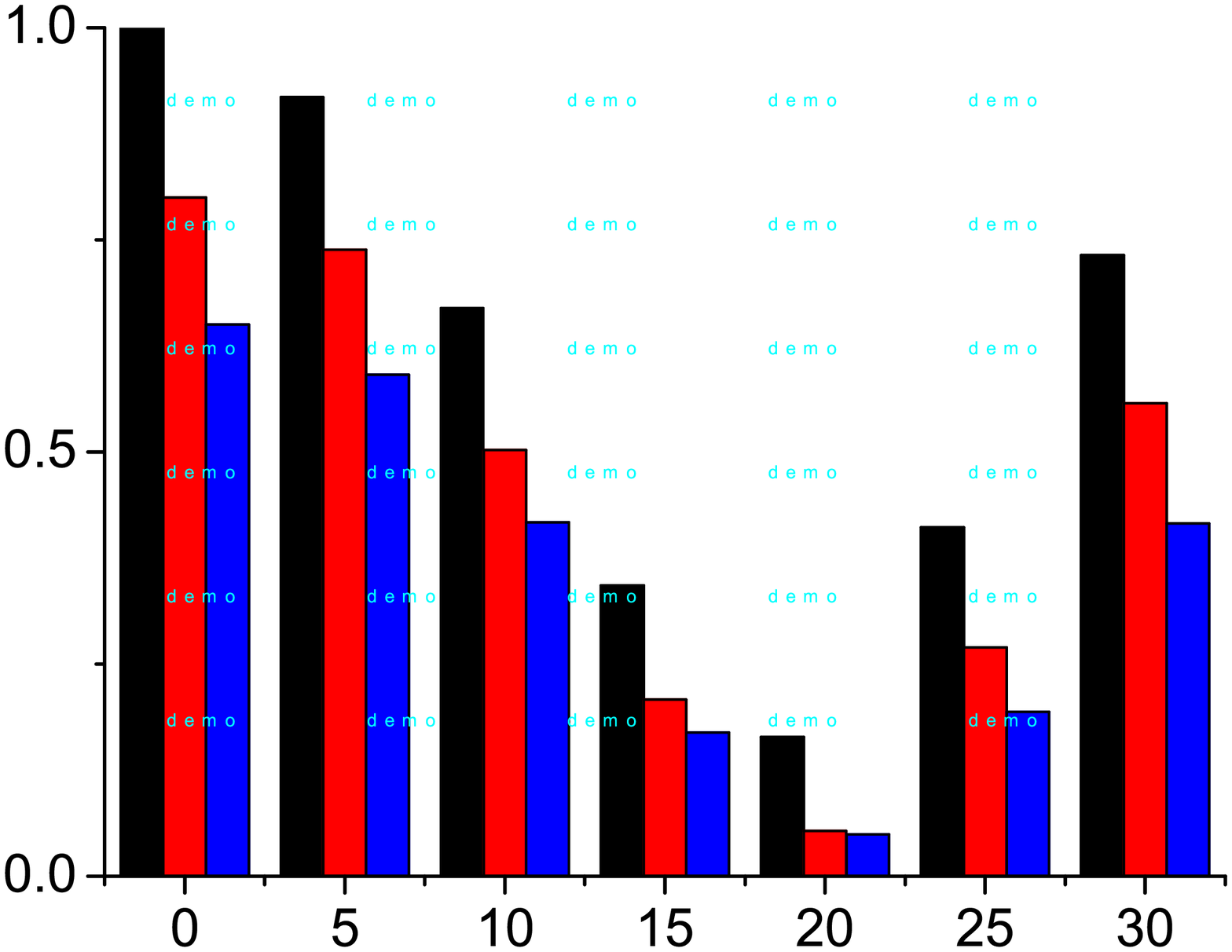}
\put(-20,120){${(b)}$}
 \put(-215,70){$\mathcal{E}(T)$}
 \put(-100,1){${n}$}
 \caption{ Degree  of entanglement
$\mathcal{E}(T)$, at the end of the rectangular pulse against the
pulse area parameter $(n)$, where only one qubit is driven by the
rectangular pulse. (a)$\Delta^{(1)}=0$, (b) $\Delta^{(1)}=1$}
\end{center}
\end{figure}

\subsection{Two driven-qubit}

\begin{figure}[t!]
\begin{center}
\includegraphics[width=18pc,height=12pc]{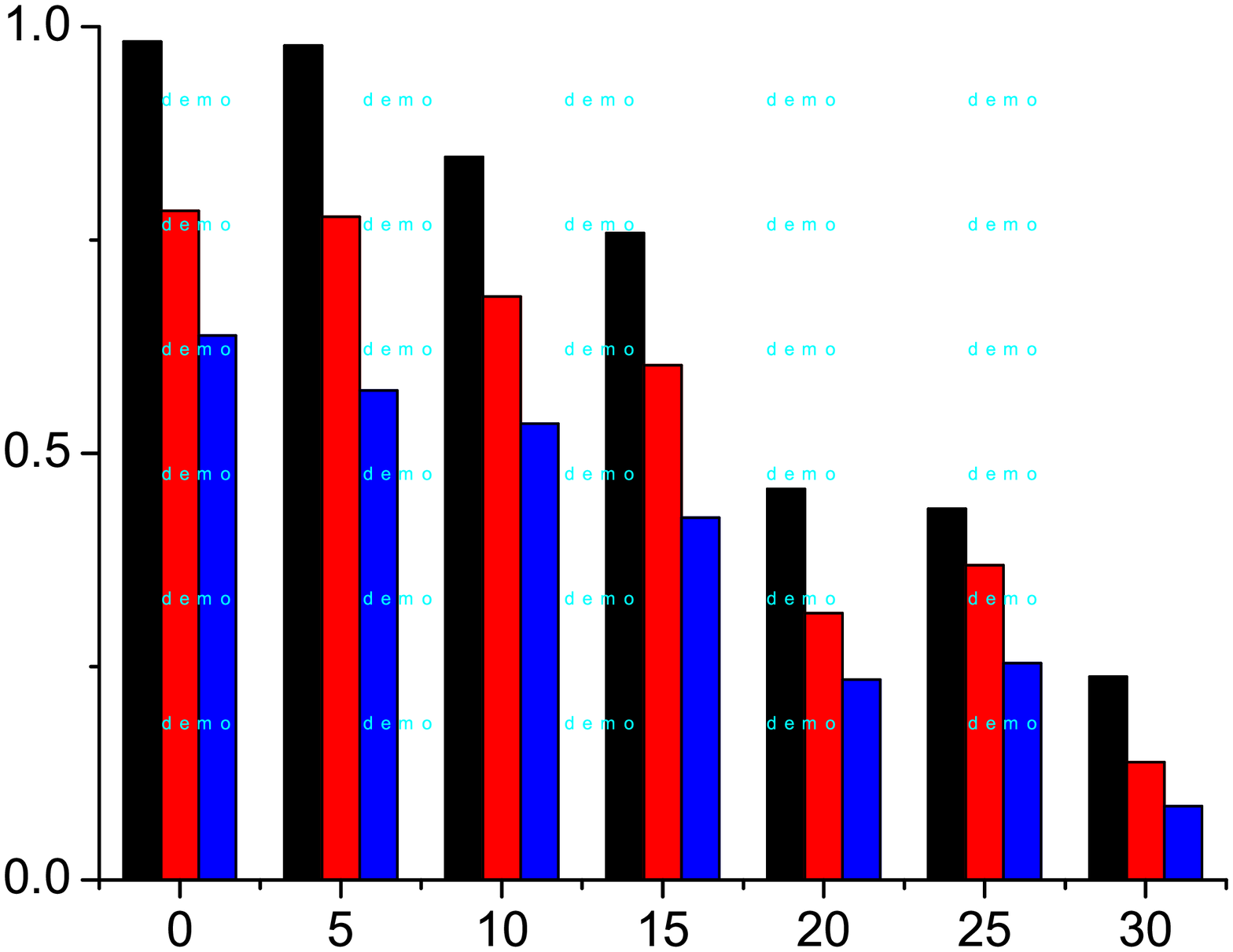}
\put(-20,120){${(a)}$}
 \put(-215,70){$\mathcal{E}(T)$}
 \put(-100,1){${n}$}
\includegraphics[width=18pc,height=12pc]{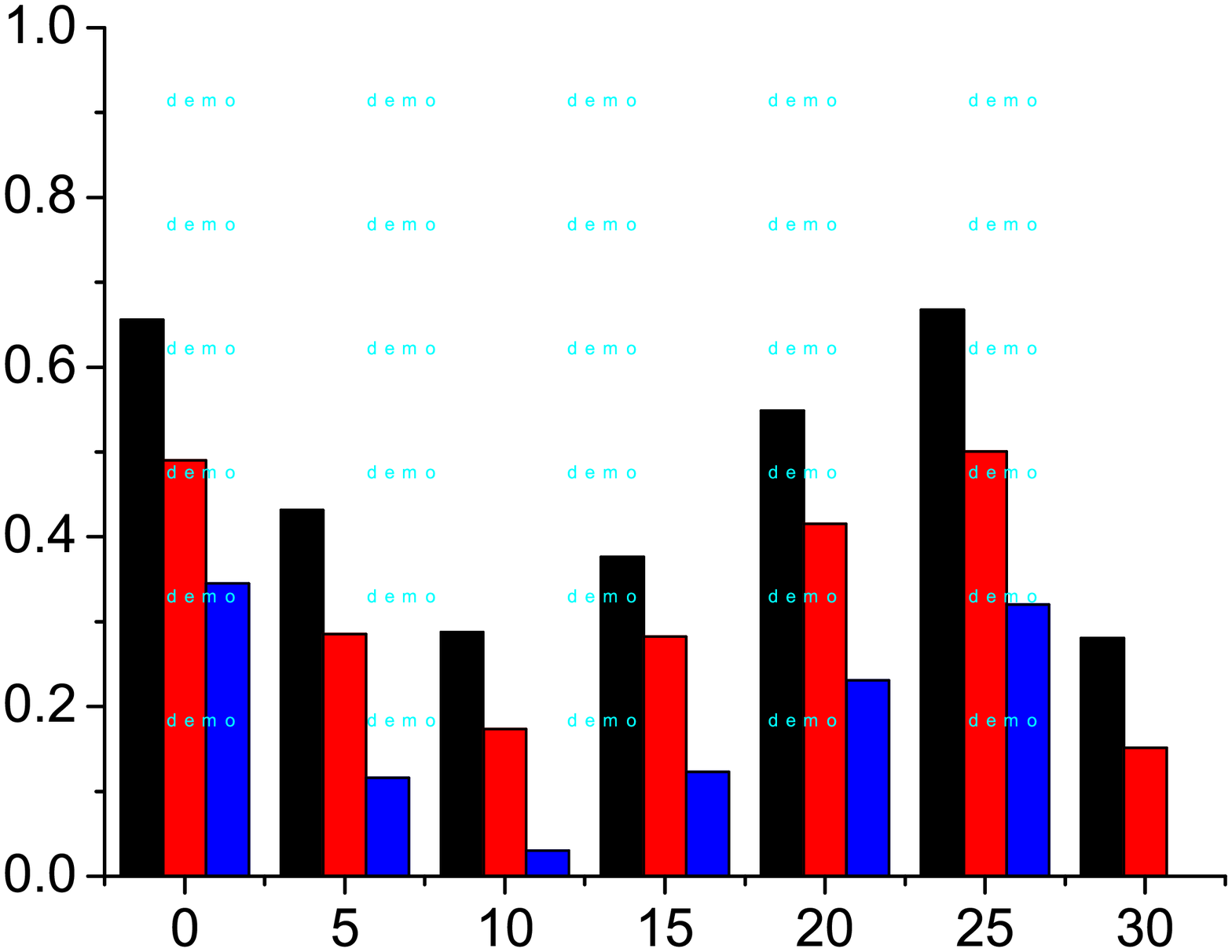}
 \put(-20,120){${(b)}$}
 \put(-215,70){$\mathcal{E}(T)$}
 \put(-100,1){${n}$}
 \caption{As Fig.(1) but with both qubits driven with rectangular
 pulses.  (a) $\Delta^{(1)}=\Delta^{(2)}=0$, (b)  $\Delta^{(1)}=\Delta^{(2)}=5$ }
\end{center}
\end{figure}

When both qubits are driven by two resonant  rectangular pulses,
Fig.(2a) shows that, $\mathcal{E}(T)$ decreases( as compared with
Fig.(1a)) as (n) increases to reach its lower bound. For
non-resonant case ($\Delta^{(1)}=\Delta^{(2)}=5$), Fig.(2b) shows
that the degree of entanglement between the two qubits decreases
to reach their lower bounds. However for larger time, the degree
of entanglement, $\mathcal{E}(T)$ increase to reach its upper
bounds. The effect of changing the Rabi frequencies induces
different oscillatory patterns.

\section{ Exponential Pulse Case}
For resonant   exponential  pulse shape as defined in (5) the
coefficients $\mathcal{C}^{(j)}_{\pm}, \mathcal{C}^{(i)}_{z}
\kappa^{(j)}_y $ and $\mathcal{D}^j_x, \mathcal{D}^{(j)}_y$ and
$\mathcal{D}_z^{(j)}$ are given by \cite{Bata1},
\begin{eqnarray}
\mathcal{C}^{(j)}_{\pm}&=&\frac{1}{2}(1\pm\cos\lambda^{(j)}(t)),
\quad \mathcal{C}^{(j)}_{z}=-i\sin{\lambda}^{(j)}(t)
\nonumber\\
D^{(j)}_x &=&0,\quad  D^{(j)}_y =\sin{\lambda}^{(j)}(t),\quad
D^{(j)}_z=\cos{\lambda}^{(j)}(t),
\end{eqnarray}
where $
\lambda^{(j)}(t)=\frac{\Omega^{(j)}}{\gamma^{(j)}_p}\Bigl(1-e^{-\gamma^{(j)}_p
t}\Bigr)$. Here, we investigate the effect of the  normalized Rabi
frequency $\Omega'=(\frac{\Omega}{\gamma_p})$ on the degree of
entanglement between the two qubits if one or both of them are
driven. Fig.(3) shows the behavior of $\mathcal{E}(T')$;
$T'=\gamma_p t$, when only one particle is driven with an
exponential pulse.  It is clear that, for small values of the
 time $T'\in[0,3]$, the entanglement $\mathcal{E}(T')$
decreases fast to reach its lower bound. Also, the initial state
with smaller degree of entanglement turns into a separable state.
However, as $T'$ increases further the degree of entanglement
between the two particles doesn't change. This means that we have
a long lived entangled state with a fixed degree of entanglement.
For larger values of Rabi frequency, $\Omega'=10$, the number of
oscillations on the time interval $T'\in[0,1.9]$ increases and the
phenomenon of sudden death and birth appear. However for larger
time, the entanglement $\mathcal{E}$ re-birthes to reach its
maximum values and then behaves invariably. All  of these
phenomena can be observed in Fig.(3b).

Fig.(4a) displays the behavior of $\mathcal{E}(T')$, when both
particles are locally pulsed with an exponential pulse.  The
attitude of the entanglement is similar to that shown in Fig.(3).
However the entanglement doesn't vanish and their lower bounds are
larger than those shown in Fig.(3), where only one particle is
driven. As the ratio $\Omega'=(\frac{\Omega}{\gamma_p})$
increases, the number of oscillations increases. As the time
increases further, the entanglement $\mathcal{E}(T')$ increases
to reach its upper bound and then behaves invariably.

\begin{figure}
\begin{center}
\includegraphics[width=16pc,height=12pc]{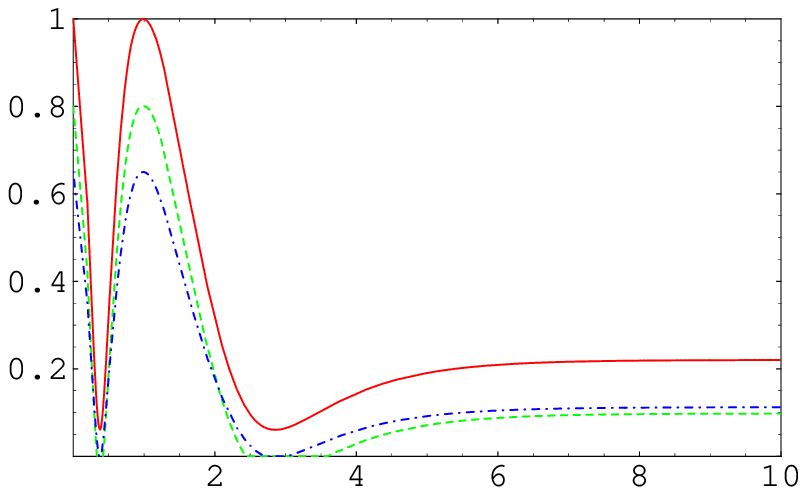}
\put(-35,127){${(a)}$}
 \put(-215,70){$\mathcal{E}(T')$}
 \put(-100,-5){${T'}$}\quad~
\includegraphics[width=16pc,height=12pc]{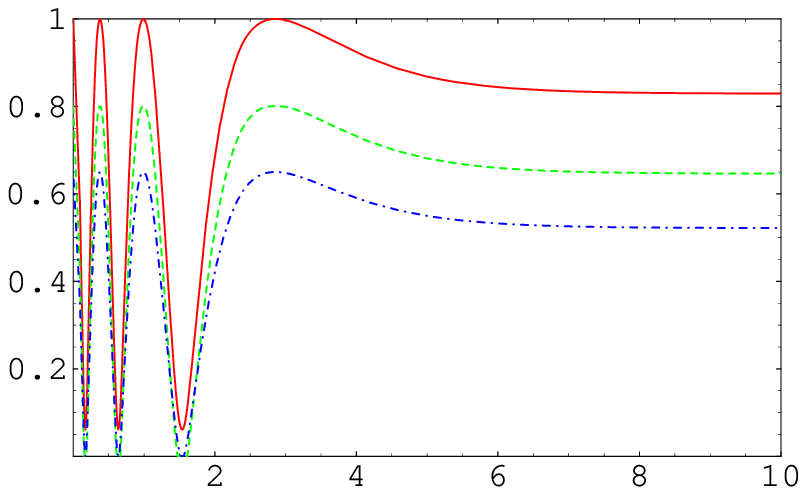}
\put(-30,127){${(b)}$}
 \put(-210,70){$\mathcal{E}(T')$}
 \put(-100,-5){${T'}$}
 \caption{Degree  of entanglement $\mathcal{E}(T')$ against the normalized time $T'=\gamma_p t$  in the  case of of local exponential
 pulse with Rabi frequency  $(\frac{\Omega}{\gamma_p})=5$(a),
 $(\frac{\Omega}{\gamma_p})=10$(b),
 where only one particle is driven. The solid curve, dot and dash-dot curves for systems
 prepared initially in maximum entangled state, partial entangled state with $c_{xx}=c_{yy}=c_{zz}=-0.9$ and
 $c_{xx}=-0.9,c_{yy}=-0.8, c_{zz}=-0.7$, respectively.}
\end{center}
\end{figure}

\begin{figure}
\begin{center}
\includegraphics[width=16pc,height=12pc]{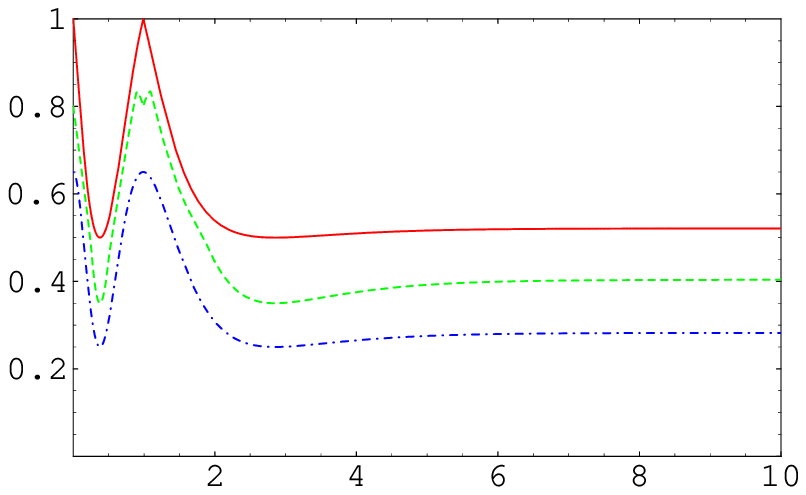}
\put(-35,127){${(a)}$} \put(-215,70){$\mathcal{E}(T')$}
 \put(-100,-5){${T'}$}\quad~
\includegraphics[width=16pc,height=12pc]{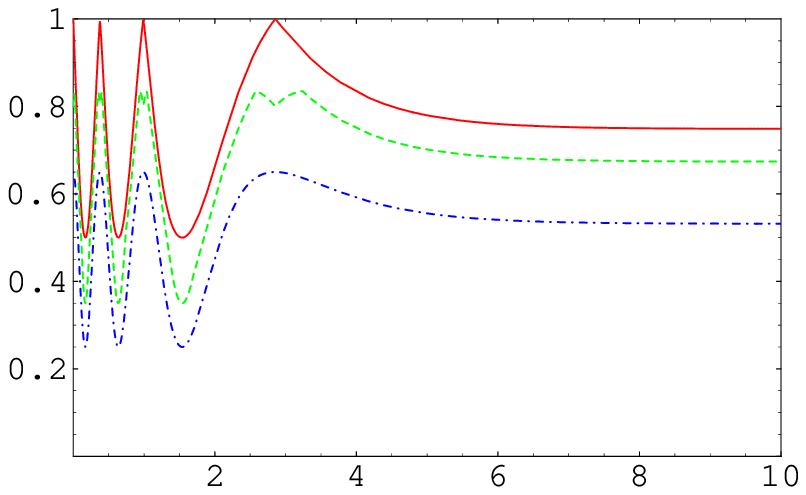}
\put(-30,127){${(b)}$}
\put(-215,70){$\mathcal{E}(T')$}
 \put(-100,-5){${T'}$}
 \caption{The same as Fig.(3) but  with both
 particles are driven with  exponential pulses.}
\end{center}
\end{figure}

\section{Combined rectangular and exponential pulse case}
In this section, we assume that Alice's qubit is driven  with a
rectangular pulse  as defined by (\ref{RectPulse}), while Bob's
qubit is driven with  exponential pulse, (\ref{ExpPulse}). The
dynamics of entanglement is displayed in Fig.(5), for different
values of the pulses parameters. Fig.(5a) describes the behavior
of the entanglement between the two qubits for $\omega^{(1)}=1$
and $(\frac{\Omega}{\gamma_p})=5$. It is clear that at $t=0$, the
entanglement  $\mathcal{E}=1$ for the  maximum  entangled state
and  $\mathcal{E}<1$ for the partially entangled states. The
general behavior is similar to that depicted for the previous
cases i.e., the degree of entanglement between the two qubits
decay as $t$ increases. However the upper and lower bounds of
$\mathcal{E}$ depends on the initial entanglement between the two
qubits. Fig.(5b) displays the behavior of $\mathcal{E}$ for larger
value of  rectangular pulse strength  i.e., we set
$\Omega^{(1)}=2$, while we fix the strength of the exponential
pulse. This behavior shows that the number of oscillations
increases and the upper and lower bounds of entanglement increase.
In Fig.(5c$\&5d$) we depict the behavior of the degree of
entanglement between the two qubits for larger values of the
exponential pulse, where we set $(\frac{\Omega}{\gamma_p})=10$. It
is clear that the upper and lower  bounds are much larger compared
with those shown in Fig.(5a$\&5b$). However for system initially
prepared in smaller degree of entanglement, the entanglement is
temporary vanishes and rebirths again \cite{Nasser2}

\begin{figure}[t!]
\begin{center}
\includegraphics[width=16pc,height=12pc]{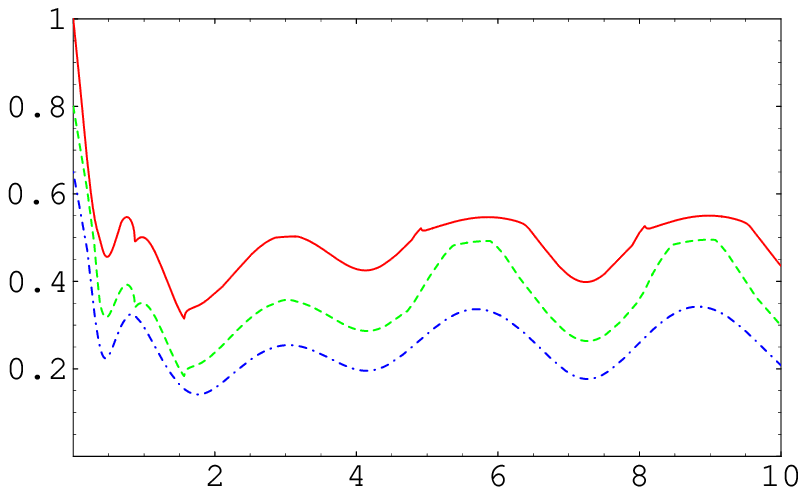}
\put(-35,127){${(a)}$}
 \put(-215,70){$\mathcal{E}(T')$}
 \put(-100,-5){${T'}$}\quad~
\includegraphics[width=16pc,height=12pc]{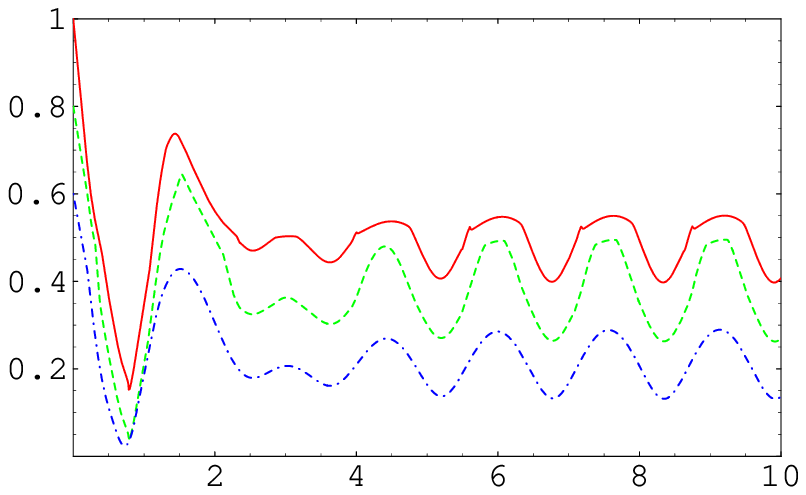}
\put(-30,127){${(b)}$} \put(-210,70){$\mathcal{E}(T')$}
 \put(-100,-5){${T'}$}\\
\includegraphics[width=16pc,height=12pc]{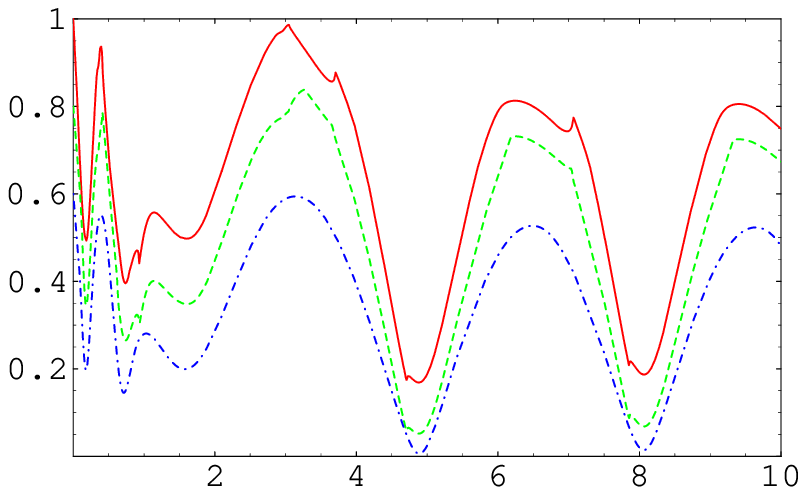}
\put(-35,127){${(c)}$} \put(-215,70){$\mathcal{E}(T')$}
 \put(-100,-5){${T'}$}\quad~
\includegraphics[width=16pc,height=12pc]{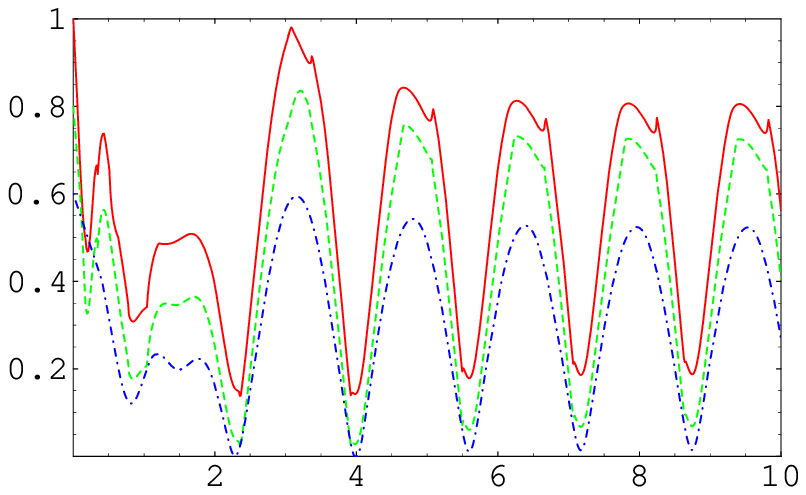}
\put(-30,127){${(d)}$}
 \put(-210,70){$\mathcal{E}(T')$}
 \put(-100,-5){${T'}$}
 \caption{Degree of  entanglement for a
system  initially prepared in maximum entangled state
(solid-curve), Werner state (dot-curve) and generalized Werner
state (dash-dot curve). It is assumed that, Alics's qubit is
driven with a rectangular  pulse and Bob's qubit is driven with an
exponential pulse. We consider the resonant case i.e.
$\Delta_1=\Delta_2=0$ and
  (a) $\Omega^{(1)}=1, (\frac{\Omega}{\gamma_p})=5$, (b) $\Omega^{(1)}=2,
  (\frac{\Omega}{\gamma_p})=5$,
  (c) $\Omega^{(1)}=1, (\frac{\Omega}{\gamma_p})=10$, (d) $\Omega^{(1)}=2, (\frac{\Omega}{\gamma_p})=10$.}
\end{center}
\end{figure}

From Fig.(5), one concludes that for small rectangular pulse area
(i.e., small $\omega^{(1)}$) the phenomena of long -lived
entanglement is remarkable. Increasing the number of oscillations
(due to large pulse area) has  a noticeable effect on the lower
and upper bounds of entanglement. For larger values of  Rabi
frequency $\omega'=\frac{\Omega}{\gamma_p}$ of the exponential
pulse the upper and lower bounds increase larger than that shown
for rectangular pulse. The phenomena of  vanishing and re-birthing
of entanglement  is shown for larger values of  $\omega'$ in the
case of  exponential pulse.

\section{Conclusion}
Effects of rectangular and exponential pulse shapes    on the
entanglement between two  initially entangled particles, maximally
or partially, are  discussed. We consider either one
 or both particles are driven.   We show that, in the presence of the rectangular
 pulse the detuning  parameter protect  the degree of entanglement
 between the two particles from vanishing, where for the resonant
 case, the entanglement vanishes for small pulse time duration.
 Consequently, one can keep on a long - lived entanglement between the two
 particles.  Although the upper bounds of the degree of entanglement between the two particles decrease as the
 detuning parameter  increases,  the lower bounds are improved.
 It is shown that, if only one particle is driven  then the
 entanglement reaches its initial values by increasing the detuning
 paprmeter. However, if both particles are driven, then the upper bounds are always smaller than the
 initial one for larger values of the detning parameter.
  Rabi- oscillations have no effect on
 the lower and upper  bounds of the survival amount of
 entanglement. However, the number of oscillations of the entanglement increases
  for larger values of  Rabi-frequency, as expected.

  In the presence of the exponential pulse, the phenomena of the
  sudden-death and -birth of entanglement appear for smaller
  time. On the other hand,  for larger time, the long-lived
  entanglement  is   invariable.  In this case, Rabi frequency plays an essential role on
  controlling the lower and upper bounds of entanglement. It is
  shown that, as one increases the  Rabi frequency the entanglement is
  protected from being lost even  for  small intervals of
  time.

For the combined case, the long-lived entanglement is depicted for
small strength values of the  rectangular and exponential pulses.
However, with larger strength of  the exponential pulse, the upper
bounds increase while the lower bounds of entanglement decreases
and completely vanish for less initially entangled pairs. The
oscillations of entanglement between its lower and upper bounds
increase as the exponential pulse increases

  {\it In conclusion:} one can protect the entanglement between
  two initially   entangled  particles by pulsed  excitation of one or both  particles.  The
  rectangular pulse keeps  a long-lived  entanglement between the two
  particles, but with variable  degree. On the other hand, excitation with the
 exponential pulse generates an  invariable long-lived entanglement.

\bigskip
{\bf Acknowledgement} The author (H A Batarfi) acknowledges the
technical and financial support of (KAU)-grant No.
(35-3-1432/HiCi).


\begin{thebibliography}{nas}

\bibitem{Barnett} S. M. Barnett," Quantum Information" (Oxford
Univ. Press, Oxford, 2009); Vlatko Vedral,"Introduction to Quantum
information  Science", (Oxford Univ. Press, Oxford, 2008).

\bibitem{Ping}P. Huang, J. Zhu, G. He and G. Zeng, J. Phys. B:At.
Mol Opt. Phys. {\bf 45} 135501 (2012); M. Siomau, J. Phys. B:At.
Mol Opt. Phys.{\bf 45} 035501 (2012).

\bibitem{Nasser0}N. Metwally Quantum Information Processing, {\bf 9} 429 (2010);
N. Metwally, J. Phys. A :Math. Theor.{\bf 44} 055305 (2011).

\bibitem{Ikram} M. I. Hussain and  M. Ikram, J. Phys. B:At.
Mol Opt. Phys. {\bf 45} 115503 (2012).

\bibitem{Gab}G. Lemos and F. Toscano, Phys. Rev. E {\bf 84}, 016220
(2011).

\bibitem {Zhao}Z. Liu and  H. Fan, Phys. Rev. A {\bf 79}, 064305
(2009).

\bibitem{Rod} P.A.Rodgers and S. Swain, Opt. Commu. {\bf 81} 291
(1991).

\bibitem{shukry0} A. Joshi and S. S. Hassan, J. Phys, B {\bf 35}
1985 (2002).
\bibitem{shukry}  S. S. Hassan, A. Joshi and N. M. M. Al-Madhari,
J. Phys. B {\bf 41} 145503 (2008); and corrigendum: J. Phys. B
{\bf 42} 089801 (2009).


\bibitem{Bata} S.S. Hassan, A. Joshi and H. A. Batarfi, Int. J.
Theor. Physics, Group theory $\&$ Nonlinear Optics, {\bf 13}
371-382( 2010).




\bibitem{Shukry1} A. S. Mohamed, S. S. Hassan and M-A. Al-Saegh,
Nonlinear Optics, Quantum Optics {\bf 36} 107 (2007).

\bibitem{shukry2}  S. S. Hassan, M. A. Al-Saegh, A. S. Mohamed and H. A.
Bararfi, Nonlinear Optics, Quantum Optics {\bf 42} 37 (2011).

\bibitem{Qader2013} M. R. Qader, J. Assoc. Arab. Univ. for
Basic $\&$ Appl. Sci. {\bf 13} 19 (2013)
\bibitem{Bata1} H. A. Batrarfi, Nonlinear Opt. Phys.$\&$ Materials,
{\bf 21} 1250025 (2012).


\bibitem{Nasser1} N. Metwally and S. S. Hassan, Nonlinear Opt. and
Quantum Optics, {\bf 44 } 267 (2012).

\bibitem{Hang} H. Shi Xu and Jing-bo-Xu, Euro. Phys. Lett, {\bf
95} 6003(2011); H. Shi Xu and Jing-bo-Xu, J. Opt. Soc. Amer. B
{\bf 29} 2074 (2012).

\bibitem{Englert} B. -G. Englert and N. Metwally, J. Mod. Opt, {\bf
47} 221 (2000); B. -G. Englert and N. Metwally, Appl. Phys. B {\bf
72} 35 (2001).
\bibitem{Nasser} N. Metwally, Int. J. Theor. Phys. {\bf 49} 1571
(2010).
\bibitem{Peres}
A. Peres, Phys. Rev. Lett. {\bf 77}, 1413 (1996); R. Horodecki, M.
Horodecki and P. Horodecki, Phys. Lett. A{\bf  222}, 1 (1996).

\bibitem{Zyc} K. Zyczkowski, P. Horodecki, A. Sanpera and M. Lewenstein, Phys. Rev. A
58, 883 (1998).





\bibitem{Nasser2} N. Metwally, M. Abdelaty and A.-S.F. Obada, Opt. Comm.  {\bf 250} 148 (2005).



\end{thebibliography}
\end{document}